\documentclass[amsmath,amssymb,11pt]{article}
\usepackage{jheppub2}
\pdfoutput=1
\usepackage{graphicx,epsfig}
\usepackage{float}
\usepackage{subfig}
\usepackage{subfloat}
\usepackage[normalem]{ulem}
\usepackage{tikz}
\usepackage{tkz-tab}

\usepackage[utf8]{inputenc}
\usepackage{hyperref}
\usepackage{caption}
\usepackage{color}
\usepackage{overpic}
\usepackage{xcolor}

\newcommand*{\affmark}[1][*]{\textsuperscript{#1}}

\definecolor{pink1}{rgb}{0.858, 0.188, 0.478}

\def\beq{\begin{equation}}
\def\eeq{\end{equation}}

\def\z{\zeta}

\title{Recent Developments in Holographic Black Hole Chemistry}

\author{Robert B. Mann,\affmark[1,2]}
\emailAdd{rbmann@uwaterloo.ca}

\affiliation{\affmark[1]Department of Physics and Astronomy, University of Waterloo,\\
 Waterloo, Ontario, N2L 3G1, Canada\\  
\affmark[2]{Perimeter Institute for Theoretical Physics, 31 Caroline St., Waterloo, Ontario, N2L 2Y5, Canada}\\
 }

\abstract{One of the major developments in classical black hole thermodynamics is the inclusion of vacuum energy in the form of thermodynamic pressure. Known as Black Hole Chemistry, 
this subdiscipline has led to the realization that anti de Sitter  black holes exhibit a broad variety of phase transitions that are essentially the same as those observed in chemical systems. Since the pressure is given in terms of 
a negative cosmological constant (which parametrizes 
the vacuum energy), the holographic interpretation of Black Hole Chemistry has remained unclear. In the last few years there has been considerable progress in developing an exact dictionary between the  bulk laws of Black Hole Chemistry and the laws of the dual Conformal Field Theory (CFT).  Holographic Black Hole Chemistry is now becoming an established subfield, with a
full thermodynamic bulk/boundary correspondence, and an emergent understanding of CFT phase behaviour  and its  correspondence in the bulk.  
Here I review these developments, highlighting key advances and briefly discussing future prospects for further research.
}
\begin{document}
\maketitle

\section{Introduction}

Black Hole Chemistry \cite{Kubiznak:2016qmn} -- an extension of black hole thermodynamics in which  the (negative) cosmological constant $\Lambda$ is identified with a (positive) thermodynamic pressure 
in $d$ spacetime dimensions 
via
\cite{Kastor:2009wy,Dolan:2011xt,Dolan:2010ha,Cvetic:2010jb,Kubiznak:2014zwa} 
\be\label{P}
P = -\frac{\Lambda}{8\pi G} = \frac{(d-1)(d-2)}{16\pi L^2 G}
\ee
-- is 15 years old, though its roots extend 
further back \cite{Teitelboim:1985dp,Creighton:1995au,Caldarelli:1999xj}.  The introduction of pressure in turn implies a conjugate thermodynamic volume $V$, 
and the mass of the black hole can be understood as thermodynamic enthalpy
\cite{Kastor:2009wy}.
Incorporation of these quantities into  black hole thermodynamics  has been shown to yield  a  range of phenomena as rich as can be found in a chemistry laboratory.
Examples include  Van der Waals   type phase transitions  for charged anti de Sitter (AdS) black holes  \cite{Kubiznak:2012wp},  reentrant phase transitions
\cite{Altamirano:2013ane,Frassino:2014pha}, triple points
\cite{Altamirano:2013uqa,Wei:2014hba,Dehghani:2020blz}, heat engines \cite{Johnson:2014yja}, polymer transitions~\cite{Dolan:2014vba}, superfluid transitions \cite{Hennigar:2016xwd}, Joule-Thompson expansions \cite{Okcu:2016tgt}, molecular microstructures \cite{Wei:2015iwa},
and most recently multicritical behaviour
\cite{Tavakoli:2022kmo,Wu:2022bdk}.  
Extensions to asymptotically flat
\cite{Wu:2022xmp}, Rindler \cite{Astorino:2016ybm,Anabalon:2018ydc,Anabalon:2018qfv}, and de Sitter 
\cite{Dolan:2013ft,Mbarek:2018bau,Simovic:2018tdy} spacetimes  have been carried out in recent years. The formalism can even be extended to solitons  \cite{Mbarek:2016mep,Quijada:2023fkc}. 

For a charged and multiply-spinning AdS black hole,  
  the   generalised Smarr relation and      first law are \cite{Dolan:2013ft}
\ba
M&=&\frac{d-2}{d-3} \bigl( 
TS+ \sum_i\Omega_i J_i \bigr)+\phi Q-\frac{2}{d-3}PV \label{smarr} \\
\delta M&=&T_H \delta S+\phi \delta Q 
+\sum_i \Omega_i \delta J_i
+ V \delta P\,,  \label{flaw}
\ea  
respectively, where 
 $M$ is the mass of the black hole whose angular momenta are  $J_i$. The   relative angular velocities ~$\Omega_i$ between the horizon and infinity  \cite{Gibbons:2004ai}
 are their   respective thermodynamic conjugates. The remaining thermodynamic variables are the  
 Hawking temperature  
 $T_H$,  the  electric charge $Q$  (with   conjugate  electrostatic potential $\phi$) and the Bekenstein-Hawking entropy $S$. In general relativity, $S = \frac{A}{4G}$, where
 the $d$-dimensional gravitational constant $G$ has 
 the same dimensions
 as the horizon area $A$ (a length dimension of $(d-2)$), 
 rendering the entropy unitless.

Despite this cornucopia of interesting results, the {\em holographic interpretation} of black hole chemistry has been somewhat  puzzling \cite{Johnson:2014yja, Dolan:2014cja, Kastor:2014dra, Zhang:2014uoa, Zhang:2015ova, Dolan:2016jjc, McCarthy:2017amh}.
Despite the fact that from a cosmological perspective $\Lambda$ 
is the energy/pressure of the vacuum, and so might be considered a variable quantity,  in 
 the AdS/CFT correspondence \cite{Maldacena:1997re,Witten:1998qj} $\Lambda$ is regarded as fixed -- it sets the asymptotic structure of the bulk spacetime and is
 proportional to the number of colours $N$ (or central charge $C$) in the dual gauge theory.
Indeed, an early version of what could be called 
holographic black hole thermodynamics argued that Anti-de Sitter (AdS) black holes were equivalent to thermal states in the dual conformal field theory (CFT)~\cite{Witten:1998zw}. Advancing this further,
the  temperature and entropy of  a Schwarzschild-AdS black hole match  the respective thermal  temperature and entropy of the dual CFT. Furthermore,   the   first-order  phase transition between 
  a large black hole and thermal AdS spacetime (known as the Hawking-Page transition \cite{Hawking:1982dh}) 
  corresponds to  the confinement/deconfinement phase transition of a quark gluon plasma~\cite{Witten:1998zw}; this transition was later shown to correspond to a liquid-solid transition in black hole chemistry \cite{Kubiznak:2014zwa}. 

The holographic dictionary
states that the thermodynamics of AdS black holes is completely equivalent to the thermodynamics of the dual CFT. 
Since a CFT is a standard unitary gauge theory (perhaps  with a large number $N$ of color degrees of freedom), the correspondence suggests that black hole evaporation is a unitary process. More generally, one expects that perplexing features of black holes can be studied in the dual field theory,
and vice-versa, via  holographic duality.

However a variable $\Lambda$ is difficult to interpret in the context of holographic duality.
Its variations correspond to changing both $N$ (or the central charge $C$)  and   the CFT volume ${\cal V}$ \cite{Johnson:2014yja,Dolan:2014cja,Kastor:2014dra,Caceres:2015vsa}; consequently the first law \eqref{flaw} cannot be straightforwardly related to the corresponding thermodynamics of the  dual field theory \cite{Karch:2015rpa, 
Sinamuli:2017rhp, Visser:2021eqk}.  Furthermore,  electric charge and its conjugate potential  both rescale with the AdS length $L$ in the correspondence.  

Recently  considerable progress 
has been
made in understanding holographic black hole chemistry. 
An exact dictionary between the  laws of Black Hole Chemistry and the  laws in the CFT has been constructed. An emergent understanding of CFT phase behaviour  and its  correspondence in the bulk is taking place, leading to
a
full thermodynamic bulk/boundary correspondence.
In this article I shall review these recent developments, highlighting some key advances and briefly discussing future avenues of research.

\section{Holographic Smarr Relation}\label{sec2}

The general assumption underlying nearly all investigations of the AdS/CFT correspondence is that the cosmological constant is a fixed parameter,
related to the number of colours $N$ in the dual gauge theory via a holographic relation of the form  \cite{Karch:2015rpa}
\begin{equation}
\label{correspond1}
\textsf{k} \frac{L^{d-2}}{16\pi G} = N^2
\end{equation}
where the numerical factor $\textsf{k}$ depends on the details of the particular holographic system
and 
 $L$  is the AdS length from  \eqref{P}.  The relation \eqref{correspond1} emerged out of  the AdS/CFT correspondence \cite{Maldacena:1997re}, in which  the near horizon geometry of $N$ coincident $D_3$ branes in type IIB supergravity corresponds to
an $\mbox{AdS}_5\times S^5$ spacetime.
Specifically, 
\begin{equation}
L^4=\frac{\sqrt{2}\ell_{Pl}^4}{\pi^2}N
\label{eq:NrhL}
\end{equation}
where $\ell_{Pl}$ is the 10-dimensional Planck length, expresses the 
correspondence between an ${\cal{N}}=4$ SU(N) Yang-Mills theory on the boundary of the $\mbox{AdS}_5\times S^5$ spacetime.

These relations suggest that variation of the AdS radius $L$ implies  variation of   the  number of colours $N$ in the corresponding Yang-Mills theory.
A consequence of this is that the variation of $\Lambda$ in the bulk corresponds to variation in the space of field theories in the boundary.
There have been a few proposals 
of this type
\cite{Johnson:2014yja,Dolan:2014cja,Kastor:2014dra}, where $V$ is interpreted in the boundary field theory as an associated chemical potential $\mu$ for colour.  

An alternative proposal is to 
 keep $N$ fixed.
The   field theory then remains unchanged, and variation of $\Lambda$ then corresponds to variation of   the curvature radius governing the space on which the field theory is defined \cite{Karch:2015rpa}.  This perspective yields a `{\it holographic Smarr relation}' based on the scaling properties of the dual field theory.  In the limit of  large $N$ the free energy ${\cal F}$ of the field theory scales  as $N^{2}$, and so
\begin{equation}\label{holosmarr}
{\cal F}(N,\mu,T,l)= N^{2}{\cal F}_{0}(\mu,T,l)  
\end{equation}
The equation of state  is
\be\label{UCFT}
E=(d-2)p{\cal V}
\ee
for a conformal field theory, and 
this can be used with \eqref{holosmarr}  to obtain the standard Smarr relation
\eqref{smarr}
(setting $J=Q=0$ for simplicity). Recalling \eqref{correspond1},   varying $\Lambda$  (or the AdS length $L$ implies
that $G$ must also be varied
since $N$ is fixed.  

This relation can be extended beyond the large $N$ limit \cite{Sinamuli:2017rhp}. The  
bulk correlates of the 
subleading $1/N$ corrections
  are related to the couplings in a class of higher curvature generalizations of
  Einstein gravity known as   
  Lovelock gravity theories.  
   In the context of string theory the additional higher curvature terms in Lovelock gravity are understood as quantum corrections to  Einstein gravity. In the context of holographic thermodynamics,  the Lovelock couplings are related  to a function of   $N$, and their variations dictate the behaviour of the corresponding CFT.

It is instructive to see how this works for the class of charged AdS black holes.  The action for Lovelock gravity minimally coupled to electromagnetism  ($F_{ab} = \nabla_{[a} A_{b]}$) is
\begin{equation}\label{LovMax}
I =\frac{1}{16\pi G }\int d^dx\sqrt{-g}\big[\sum^{\frac{d-1}{2}}_{k=0}\hat{\alpha}_{(k)}L^{(k)}-4\pi G F_{ab}F^{ab}\big].
\end{equation}
where $\hat{\alpha}_{(k)}$  is the Lovelock coupling constant for the $k$-th power of curvature, and 
\begin{equation}
\label{lagrangian2}
L^{(k)}=\frac{1}{2^k}\delta^{a_1b_1...a_kb_k}_{c_1d_1...c_kd_k}R_{a_1b_1}^{c_1d_1}...R_{a_kb_k}^{c_kd_k}
\end{equation}
is the Euler density of dimension $2k$ composed of powers of the Riemann tensor
$R_{a_kb_k}^{c_kd_k}$,
with $\delta^{a_1b_1...a_kb_k}_{c_1d_1...c_kd_k}$   totally antisymmetric in both sets of indices of the Kronecker delta functions.  Variation of the metric and gauge field yield the equations of motion 
\begin{equation}
\label{einsteinmaxwell}
\sum^{\frac{d-1}{2}}_{k=0}\hat{\alpha}_{(k)}{G^{(k)}}^{a}_{b}=8\pi G \big[F_{ac}F_b^c-\frac{1}{4}g_{ab}F_{cd}F^{cd}\big]
\qquad
\nabla_a F^{a b} = 0
\end{equation}
where
\begin{equation}
{G^{(k)}}^{a}_{b}=-\frac{1}{2^{k+1}}\delta^{aa_1b_1...a_kb_k}_{bc_1d_1...c_kd_k}R_{a_1b_1}^{c_1d_1}...R_{a_kb_k}^{c_kd_k}
\end{equation}
each of which independently satisfy 
\be\nabla_a {G^{(k)}}^{a}_{b}=0
\ee
which is a generalization of the Bianchi identities.

Imposing  spherical symmetry 
\begin{eqnarray}
\label{sphericalsym}
ds^2 = -f(r)dt^2+f(r)^{-1}dr^2+r^2d\Omega^{2}_{(\kappa)d-2} \qquad 
F = \frac{Q}{r^{d-2}}dt\wedge dr
\end{eqnarray}
where $d\Omega^{2}_{(\kappa)d-2}$ is the line element of a $(d-2)$-dimensional compact space of constant curvature $(d-2)(d-3)\kappa$
($\kappa=-1,0,1$),  the equations of motion (\ref{einsteinmaxwell}) become \cite{Frassino:2014pha}
\begin{eqnarray}
\sum^{\frac{d-1}{2}}_{k=0}\alpha_k\bigg(\frac{\kappa-f}{r^2}\bigg)^k&=&\frac{16\pi G M}{(d-2)\omega^{\kappa}_{d-2}r^{d-1}}-\frac{8\pi G Q^2}{(d-2)(d-3)r^{2(d-2)}} 
\label{f-def}
\end{eqnarray}
where $\omega^{(1)}_{d-2}=\frac{2\pi^{(d-1)/2}}{\Gamma((d-1)/2)}$~~ and 
\begin{eqnarray}
\alpha_0=\frac{\hat{\alpha}_{(0)}}{(d-1)(d-2)},~~~~~~~~\alpha_1=\hat{\alpha}_{(1)} \qquad 
\alpha_k= \hat{\alpha}_{(k)}\prod^{2k}_{n=3}(d-n)~~~~~~ \mbox{for}~~k\geq 2
\end{eqnarray}
is a  useful rescaling of the Lovelock couplings.

 Without solving 
\eqref{f-def}, 
the first law of thermodynamics and the Smarr relation 
can respectively be shown to be  \cite{Kastor:2010gq}
\begin{eqnarray}\label{flawLove}
\delta M&=&T\delta S 
+\phi\delta Q-\frac{1}{16\pi G}\sum^{\frac{d-1}{2}}_{k=0}\Psi^{(k)}\delta\hat{\alpha}_{(k)} \\
(d-3)M&=&(d-2)TS+(d-3)\phi Q + \sum^{\frac{d-1}{2}}_{k=0}\frac{2(k-1)}{16\pi G}\Psi^{(k)}\hat{\alpha}_{(k)}
\label{smarrlov}
\end{eqnarray}
where 
\begin{eqnarray}
M&=&\frac{\omega^{(\kappa)}_{d-2}(d-2)}{16\pi G}\sum_{k=0}\alpha_k\kappa^k{r_+}^{d-1-2k}+\frac{\omega^{(\kappa)}_{d-2}Q^2}{2(d-3){r_+}^{d-3}}\nonumber\\
Q&=&\frac{1}{2\omega^{(\kappa)}_{d-2}}\int\ast F  \qquad \phi = \frac{\omega^{(\kappa)}_{d-2}Q}{(d-3)r_+^{d-3}}
\nonumber\\
T&=&\frac{1}{4\pi r_+D(r_+)}\bigg[\sum_{k=0}\kappa\alpha_k(d-2k-1)(\frac{\kappa}{r_+^2})^{k-1}- \frac{8\pi GQ^2}{(d-2)r_+^{2(d-3)}}\bigg] 
\label{thermoquantities} \\
S&=&\frac{\omega^{(\kappa)}_{d-2}(d-2)}{4G}\sum_{k=0}\frac{k\kappa^{k-1}\alpha_kr_+^{d-2k}}{d-2k} \qquad D(r_+)\equiv \sum_{k=1}k\alpha_k(\kappa r_+^{-2})^{k-1} \nonumber
\end{eqnarray}
are the respective mass $M$, charge $Q$ (with conjugate potential $\phi$) and  (if it is a black hole)  temperature $T$ , and entropy $S$.  The quantities
\begin{equation}
\label{conjugate}
\Psi^{(k)}=\frac{ \kappa^{k-1}\omega^{(\kappa)}_{d-2}(d-2)}{16\pi G}r_+^{d-2k}\bigg[\frac{\kappa}{r_+}-\frac{4\pi kT}{d-2k}\bigg], ~~~~~k\geq 0
\end{equation}
are the thermodynamic conjugates to the  $\hat{\alpha}_{(k)}$; both can be regarded as thermodynamic variables \cite{Kastor:2010gq}. 
The horizon radius $r_+$  is the largest root of $f(r)=0$, and 
 the thermodynamic pressure and  volume are given by the $k=0$ terms
\begin{eqnarray}\label{PandV}
P&=&-\frac{\Lambda}{8\pi G} = \frac{(d-1)(d-2)}{16\pi G}\alpha_0 \qquad
V =\omega^{(\kappa)}_{d-2} ~\frac{r_+^{d-1}}{d-1}
\end{eqnarray}
so that $P_bV_b=\alpha_0\Psi^{(0)}$.  
The CFT volume is    $v=\omega_n^{(\kappa)}R^n$
where $R$ is the radius of the sphere on which the CFT is defined. The CFT pressure
 $p$ has a length dimension of $-(n+1)$.

In Lovelock gravity the 
free energy does not directly scale with the number of colours $N$, but rather has the form
\cite{Karch:2015rpa,Sinamuli:2017rhp}
\begin{equation}\label{felov}
{\cal F}(N, \mu, T, \alpha_j,  R)=\sum_{k=0}g_k(N){\cal F}^k(\mu, T, \alpha_j,  R)
\end{equation}
where the $g_k(N)$ are assumed to be polynomial functions, with $g_0(N) = N^2$, where $N^{2}$ is the central charge. The thermal properties of AdS black holes  can be reinterpreted as those of a CFT  at the same finite temperature according to
 the AdS/CFT correspondence  \cite{Witten:1998zw}, and so  
\begin{eqnarray}\label{Omega-eq}
{\cal F} = M-TS-\phi Q &\leftrightarrow & 
\tilde{{\cal F}}=\tilde{M}-T\tilde{S}-\tilde{\phi}\tilde{Q}
\end{eqnarray}
for the grand canonical free energy. 
where   $\tilde{M}, \tilde{S}$, $\tilde{Q} = Q L/\sqrt{G}$, $\tilde{\phi} = \phi\sqrt{G}/L$
are the respective mass,   entropy, charge, and electrostatic potential per unit volume of the CFT, along with
$\alpha_k^F=\alpha_k \left(\frac{L^2}{G}\right)^{(1-k)}$ and $\Psi_F^{(k)}=\Psi^{(k)}\left(\frac{L^2}{G}\right)^{(k-1)}$, with other quantities unchanged.

 The grand canonical free energy \eqref{felov}   is a polynomial in $1/N^2$. 
 Only the leading term (of order $N^2$)
 in this polynomial is taken into account 
 in Einstein gravity   \cite{Karch:2015rpa}, since  the large $N$ limit is taken.
However additional powers of  $1/N^2$
need to be included in
the Lovelock case   because of contributions from  higher curvature terms. 

These can be inferred from the  dimensionality of the couplings.  For some length scale $l$
\begin{equation}
\label{extendedholo2}
\alpha_k\sim l^{2(k-1)} ~~~~~~~\mbox{or}~~~~~
[\alpha_k]=2(k-1)
\end{equation}
implying
\begin{equation}
\label{extendedholo1}
g_k(N) = \beta_k(\alpha_k)^{\frac{d-2}{2(k-1)}}
\end{equation}
  which  is  
\begin{equation}\label{k=0}
N^2 = \beta_0 L^{d-2}
=\frac{\delta L^{d-2}}{16\pi G}
\end{equation}
for   $k=0$, 
recovering the relationship \eqref{correspond1}  \cite{Karch:2015rpa} where $\delta$ is an arbitrary dimensionless constant. 
 
For  any arbitrary function $X$ of the parameters $\alpha_k$,
(\ref{extendedholo1}) implies
\begin{equation}
\label{holosmarr1}
2(k-1)\alpha_k\frac{\partial X}{\partial\alpha_k}= (d-2) g_k \frac{\partial X}{\partial g_k}
\end{equation}
for any arbitrary function $X$ of the parameters $\alpha_k$. Since 
$-2\alpha_0\partial_{\alpha_0} =L\partial_L$ (from \eqref{P} and \eqref{PandV}), this relation can be written as  
\begin{equation}
\label{homo3}
L\frac{\partial}{\partial L}+\sum_{k=1}2(k-1)\alpha_k\frac{\partial}{\partial{\alpha_k}}=(d-2)\sum_{k=0}g_k\frac{\partial}{\partial g_k}
\end{equation}
after summing over $k$. In particular
\begin{equation}
\quad 2\sum_{k=0}(k-1)\alpha_k\Psi^{(k)}=(d-2)\sum_{k=0}g_k\frac{\partial{\cal F}}{\partial g_k} =  (d-2){\cal F}
\label{Euler1}
\end{equation}
using the Euler scaling relation $f(tx_1,\ldots,tx_m)=t^n f(x_1,\ldots, x_m)$ for a  homogeneous function of order $n$, 
where $\Psi^{(k)} = \frac{\partial{\cal F}}{\partial{\alpha_k}}$ and 
\begin{equation}
{\cal F} =\sum_{k=0}g_k\frac{\partial{\cal F}}{\partial g_k}
\end{equation} 
since ${\cal F}$ is an homogeneous function of the $g_k$ of degree 1.

Of course the free energy is not only a function of the $g_k$, but also depends on 
$R$ and the charge(s).   
The derivative of a function $f(L, Z)$ with respect to $L$ 
is
\begin{equation}
\partial_l f(l, Z)|_{Z_b}=\partial_lf|_Z+ z \frac{Z}{l}\partial_Z f|_l.
\end{equation}
if $Z$ scales as $Z=Z_0 L^z$ for some constant $Z_0$. For a charged AdS black hole  
\begin{equation}
A_b=LA, ~~~~\phi_b=L\phi, ~~~~~Q_b=Q/L \qquad R= R_0 L
\end{equation}
upon converting to a canonical normalized field strength of dimension $2$, where $A_b$ is the horizon area.  The scaling of $R$ is a consequence of the form
\begin{equation}
ds^{2}_{boundary}=-dt^2+l^2d\Omega^{2}_{d-2}
\end{equation}
of the boundary metric \cite{Karch:2015rpa}.   

Consequently \eqref{Euler1} becomes
\begin{eqnarray}
\sum_{k=0}2(k-1)\alpha_k\Psi^k
&&=(d-2)\sum_{k=0}g_k\partial_{g_k}{\cal F}\big|_{\phi, T}+R\partial_R{\cal F}\big|_{\phi, T,\alpha_{k\geq 1}} + Q\partial_Q{\cal F}\big|_{\phi, T,\alpha_k}\nonumber\\
&&=(d-2){\cal F} - M-\phi Q\nonumber\\
&&=(d-3)M-(d-2)TS-(d-3)\phi Q 
\label{smarr2}
\end{eqnarray}
recovering the bulk Smarr relation \eqref{smarr} 
upon using \eqref{Omega-eq}.  

The remaining task is to see how the $g_k(N)$ can be approximated. Since we expect terms for $k\geq 1$ to be suppressed relative to the $k=0$ case \eqref{k=0}, the   ansatz    
 \begin{equation}
\label{correspondance}
g_k(N)\equiv {\cal{ O}}(N^{2(1-k)}) 
\end{equation}
will be employed.
In the context of  higher curvature theories of gravity, the additional contributions   can be understood as correction terms to the Einstein-Hilbert action, or in other words, we expect
\begin{eqnarray}
&& L^{(0)}\sim {\cal{R}}^0\rightarrow N^2 \qquad L^{(1)}\sim {\cal{R}}^1 \rightarrow   N^0 \qquad L^{(2)}\sim {\cal{R}}^2 \rightarrow   N^{-2} \nonumber\\
&& \qquad \cdots \quad 
L^{(k)}\sim {\cal{R}}^k \rightarrow  N^{2(1-k)}
\quad 
\cdots 
\end{eqnarray}
with ${\cal{R}}$ some scalar measure of the curvature.

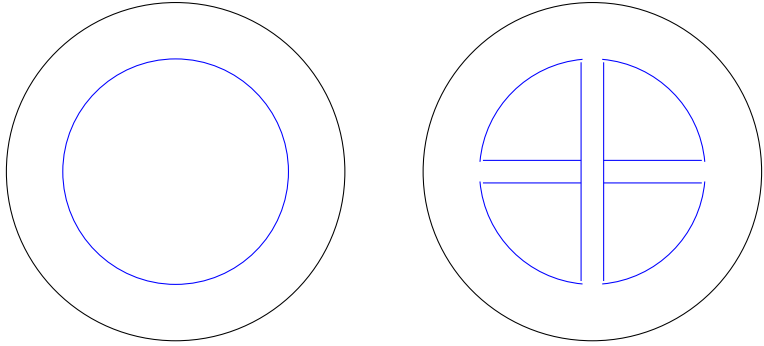
\begin{figure}
\centering
\begin{tikzpicture}[scale=1.5]
\draw[blue] (0,0) circle (1);
\draw (0,0) circle (1.5);
\end{tikzpicture}
$\qquad$
\begin{tikzpicture}[scale=1.5]
\draw (0,0) circle (1.5);
\draw[blue] (5:1) arc (5:85:1);
\draw[blue] (95:1) arc (95:175:1);
\draw[blue] (185:1) arc (185:265:1);
\draw[blue] (275:1) arc (275:355:1);
\draw[blue] (0.1,0.97)--(0.1,-0.97);
\draw[blue] (-0.1,0.97)--(-0.1,-0.97);
\draw[blue] (-0.97,-0.1)--(-0.1,-0.1);
\draw[blue](0.1,-0.1)--(0.97,-0.1);
\draw[blue] (0.1,0.1)--(0.97,0.1);
\draw[blue] (-0.1,0.1)--(-0.97,0.1);
\end{tikzpicture}
\caption{The contribution to the
scattering amplitude of the
planar diagram on the left is proportional to $N^2(g_{YM})^0=N^2\lambda^0 = N^2$ 
and corresponds to $L^{(0)}$ in the Lovelock theory.  Contributions to the scattering amplitude from $L^{(1)}$ yield non-planar diagrams of the form shown on the right, and are  proportional to $N^2 (g_{YM})^4=N^0\lambda^2 = \lambda^2$.   The Yang-Mills coupling $g_{YM}$ appears at each vertex of the diagram. This figure originally appeared in ref.~\cite{Sinamuli:2017rhp} and is reprinted with permission.}
\label{figure1}
\end{figure}
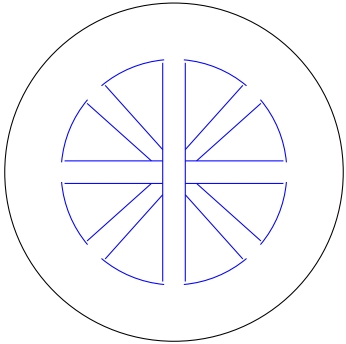
\begin{figure}
\centering
\begin{tikzpicture}[scale=1.5]
\draw (0,0) circle (1.5);
\draw[blue] (5:1) arc (5:40:1);
\draw[blue] (50:1) arc (50:85:1);
\draw[blue] (95:1) arc (95:130:1);
\draw[blue] (140:1)arc (140:175:1);
\draw[blue] (185:1) arc (185:220:1);
\draw[blue] (230:1)arc (230:265:1);
\draw[blue] (275:1) arc (275:310:1);
\draw[blue] (320:1) arc (320:355:1);
\draw[blue] (0.1,0.97)--(0.1,-0.97);
\draw[blue] (-0.1,0.97)--(-0.1,-0.97);
\draw[blue] (-0.97,-0.1)--(-0.1,-0.1);
\draw[blue] (0.1,-0.1)--(0.97,-0.1);
\draw[blue] (0.1,0.1)--(0.97,0.1);
\draw[blue] (-0.1,0.1)--(-0.97,0.1);
\draw[blue] (0.77,0.61)--(0.2,0.1);
\draw[blue] (-0.1,-0.2)--(-0.61,-0.77);
\draw[blue] (-0.2,0.1)--(-0.77,0.61);
\draw[blue] (0.1,-0.2)--(0.61,-0.77);
\draw[blue] (0.1,0.2)--(0.61,0.77);
\draw[blue] (-0.1,0.2)--(-0.61,0.77);
\draw[blue] (-0.2,-0.1)--(-0.77,-0.61);
\draw[blue] (0.2,-0.1)--(0.77,-0.61);
\end{tikzpicture}
\caption{This diagram can be understood as two overlaid copies of the 4-vertex non planar diagram 
in figure~\ref{figure1}. Non-planar diagrams of this form yield contributions to the scattering amplitude proportional to $N^2 (g_{YM})^8=N^{-2}\lambda^4$,  and result from $L^{(2)}$ terms in the Lovelock theory. This figure originally appeared in ref.~\cite{Sinamuli:2017rhp} and is reprinted with permission. }
\label{figure2}
\end{figure}

These terms make distinct contributions to the scattering amplitude.  To illustrate this, 
 consider    a field theory coupled to a Yang-Mills gauge theory,
 whose  Lagrangian $L^{(1)}$ gives rise to planar and non planar diagrams.    Gauge field self-interaction terms arise from
 $L^{(1)}$, whereas  higher order diagrams come explicitly from  higher curvature terms ($R_{d+1}=R_d+gF^2+...$); these will yield  non-planar diagrams
 as shown in figures~\ref{figure1} and~\ref{figure2}. Non-planar diagrams with 4 vertices contribute terms proportional to $N^2g_{YM}^4=\lambda^2$ 
 -- of order $N^0$ -- 
 to the scattering amplitude, 
 where $g_{YM}$ and $\lambda$ are the respective Yang-Mills and 't Hooft couplings. For $k=2$ non-planar diagrams of the form shown in figure~\ref{figure2} will contribute  
terms proportional to $N^2 (g_{YM})^8=N^{-2}\lambda^4$. The $L^{(k)}$ term in the Lovelock action will yield  diagrams having a stack of $k$ copies of the four vertices non planar diagram, whose the contribution to the scattering amplitude is proportional to $N^2 (g_{YM})^{2k}=N^{2(1-k)}\lambda^{2k}$, commensurate with  the correspondence
\eqref{correspondance}. These results are applicable to any higher-curvature theory of gravity.

\section{Variable $G$}
\label{sec3}

One approach in understanding
the variation of $L$ whilst retaining the consistency of the relation
\eqref{correspond1} is to vary the gravitational constant $G$ \cite{Cong:2021fnf}.
This has profound consequences for   black hole chemistry and its holographic interpretation.

Recall that   the first law is 
\be\label{FirstHol}
\delta E=T\delta S-p d{\cal V}+\tilde\phi \delta \tilde Q+\Omega \delta J+ \tilde \mu \delta C \,, 
\ee
for the CFT \cite{Visser:2021eqk}, 
where   $p$ and ${\cal V}={\cal V}_0 L^{d-2}$ are the CFT pressure and volume, taking the radius of the sphere on which the CFT is defined to be the AdS length.  The 
quantity $E$
is the CFT energy; it is not the enthalpy.  The angular momentum and conjugate angular velocity 
are respectively $J$ and $\Omega$, and $ \tilde Q, \tilde\phi$ are the respective holographic charge and conjugate potential. 
The quantity $\tilde \mu$ is the chemical potential for the central charge $C$.
For $SU(N)$ gauge theories with conformal symmetry
$C\propto N^2$; more generally 
$C$ is proportional to a power of $N$.

Dimensionally these quantities scale as
\ba
[E]&=&[T]=[\Omega]=[\tilde \mu]=\frac{1}{L}\, \quad [{\cal V}]=L^{d-2}\, 
\quad {}[S] =[\tilde Q]=[J]=[C]=L^0\,
\label{units1}
\ea
which in turn implies \eqref{UCFT} 
from  the standard    Euler scaling argument.  

Using the fact that $S,\tilde Q,J, C$ scale as $C$, but ${\cal V}$ does not \cite{Visser:2021eqk} yields 
the {\em holographic Smarr}   relation  \cite{Visser:2021eqk}
\be\label{SmarrHol}
E=TS+\tilde \phi \tilde Q+\Omega J+\tilde\mu C\,, 
\ee
via 
a similar Eulerian scaling argument.  
Note that   $d$-dependent factors 
are absent in 
\eqref{SmarrHol} 
and that  no $p-{\cal V}$ term appears.  
Noting the relationship between $C$ and $N$, equation
\eqref{correspond1} 
can be written as
\be\label{Ck}
C= \textsf{k} \frac{L^{d-2}}{16\pi G}\,
\ee
and furthermore
\be\label{Qt}
E=M\,,  \qquad \tilde Q= \frac{QL}{\sqrt{G}}\,,\qquad \tilde \phi = \frac{\phi\sqrt{G}}{L}\,,
\ee
can likewise be identified.  

Inserting these relations into \eqref{FirstHol} 
yields
\begin{align}
\delta (GM) & =\frac{\kappa}{8\pi } \delta A+\Omega\; \delta( GJ)+ \sqrt{G} {\phi}\; \delta( \sqrt{G} Q) -\frac{V}{8\pi} \delta \Lambda  \nonumber\\
 \Leftrightarrow \quad \delta M & =\frac{\kappa}{8\pi G} \delta A+\Omega \delta J+\phi \delta Q -\frac{V}{8\pi G} \delta \Lambda-\alpha \frac{\delta G}{G}\,, 
 \label{Bulk1}
\end{align}
using  \eqref{P} 
and \eqref{SmarrHol}, 
where
\ba 
\alpha&=&\frac{1}{2}\phi Q+\tilde \mu C+TS=M-\Omega J-\frac{1}{2}\phi Q\,,
\label{alp}
\\
V&=&\frac{8\pi G l^2}{(d-1)(d-2)}\Bigl(M-\phi Q-(d-2)C\tilde \mu\Bigr)\,\nonumber\\
&=&\frac{(d-3)}{2P}\Bigl(\frac{d-2}{d-3}(\Omega J+TS)+\phi Q-M\Bigr)\,.
\label{Vsmarr}
\ea
The last relation is   the bulk Smarr relation \eqref{smarr} (with vanishing Lovelock couplings).

Allowing $G$ to vary raises it to the status of a thermodynamic variable, whose conjugate is  
 $\alpha/G$.  When  $G$ is held fixed, the first law \eqref{Bulk1} reduces to \eqref{flaw}.  Dimensional analysis of the various quantities  implies 
\begin{align}\label{units2} 
&[\phi] =  L^{\frac{2-d}{2}}\,, \quad [Q]=L^\frac{d-4}{2} \,,
\quad [A]=[G]=L^{d-2}\,, 
\\{ }
& [M]= \frac{1}{L}\,,\quad  
[\Lambda] = \frac{1}{L^2}\,, \quad [P]=\frac{1}{L^d}\,, \quad
[V]= L^{d-1}\,,  \nonumber
\end{align} 
and yields  
the bulk Smarr relation \eqref{smarr} provided \eqref{alp}  holds.  Since this relation follows from the holographic Smarr relation, we see 
a connection between the two.
 
We can use \eqref{Ck} to write variations of $G$ in terms of $C$ and $L$ (that is, $P$), and in turn
rewrite the bulk first law  
\eqref{Bulk1}  as
\be
\delta M = T \delta S+\Omega \delta J+\phi \delta Q+V_C \delta P+\mu \delta C\,, \label{FirstC} 
\ee
where
\be\label{VC}
V_C =\frac{2M+(d-4)\phi Q}{2 d P}\,,\quad  \mu =\frac{2P(V_C-V)}{C(d-2)}\,,
\ee
are  a new thermodynamic volume $V_C$ and chemical potential  $\mu$.

\section{Variable Conformal Factor} \label{sec4}

The variation of $G$ in the previous section kept the variations of $C$ and ${\cal V}$ independent.  Since ${\cal V}={\cal V}_0 L^{d-2}$, variations of the cosmological constant (the bulk pressure) induce   variations of the CFT volume ${\cal V}$, and without variation of $G$, the corresponding CFT first law would be  degenerate, since   the $\mu \delta C$ and $-p \delta {\cal V}$ would not be independent, leaving  the CFT interpretation of   black hole chemistry obscure at the least. 
However variation of $G$ as a thermodynamic quantity is likewise fraught with unclear interpretation.  For this reason another approach is desirable.

Fortunately such an approach has recently been developed \cite{Ahmed:2023snm}.  The boundary metric of the dual CFT is actually
\cite{Gubser:1998bc,Witten:1998qj}
\be
ds^2= \omega^2 \Bigl(-dt^2+L^2
d\Omega^{2}_{(\kappa)d-2}
\Bigr)
\ee
and is obtained by the conformal completion of the bulk AdS spacetime,  
where   $\omega$ is an `arbitrary'  dimensionless  conformal factor that is a function of boundary coordinates, reflecting the conformal symmetry of the boundary theory.   

If $\omega$ is set to $1$
(the standard choice) than   the CFT volume ${\cal V} \propto L^{d-2}$, as in the previous section.
However we saw in section~\ref{sec2}
that in general the radius of the space on which the CFT is defined need not be given in terms of $L$. 
This suggests that  $\omega$ be treated as a (dimensionless) thermodynamic parameter, rather than a function of the boundary coordinates, making
 the central charge and volume independent variables.  
    This is not without precedent.
For the $\kappa=0$ planar AdS black brane case, variations of   volume $\mathcal{V}$ and central charge $C$ are  clearly   independent; varying the former corresponds to changing the number of points in the system, whereas varying the latter corresponds to varying the number of degrees of freedom at each point. Since the planar case can be reached as a limit of the $\kappa=1$ spherical case, it is reasonable to expect this independence to extend to non-planar cases.

Regarding $\omega$ as another variable is tantamount to changing the CFT volume, which  is now proportional to
\be \label{volomeg}
{\cal V} \propto (\omega L)^{d-2}\,. 
\ee
and for Einstein-Maxwell theory implies the following generalized dictionary 
\ba \label{xdictionary}
\tilde S&=& S=\frac{A}{4 G}\,,\quad 
\tilde E=\frac{M}{\omega}\,,\quad \tilde T=\frac{T}{\omega}\,,\quad \tilde\Omega=\frac{\Omega}{\omega}\,,\nonumber\\
\tilde J &=& J, \quad \tilde \Phi=\frac{\Phi\sqrt{G}}{\omega L}\,,\quad \tilde Q=\frac{QL}{\sqrt{G}}\,.
\ea
between the bulk (no tildes) and dual CFT (with tildes) thermodynamic quantities.  Note that 
variations in $L$ (or $\Lambda$)
induce variations in $C$ via \eqref{Ck} with $G$ fixed whilst keeping ${\cal V}$ independent, and likewise enter in the  variation of  spatial volume  and electric charge.   
 
Using the Smarr relation \eqref{Vsmarr}, the  first law \eqref{Bulk1} (for fixed $G$)  can be rewritten as 
\begin{align}\label{Eq8}
\delta \Bigl(\frac{M}{\omega}\Bigr)
&= \frac{T}{\omega}\delta \Bigl(\frac{A}{4 G}\Bigr)+\frac{\Omega}{\omega}\delta J+\frac{\Phi\sqrt{G}}{\omega L}\delta \Bigl(\frac{QL}{\sqrt{G}}\Bigl)\nonumber\\
&\qquad +   \Bigl(\frac{M}{\omega}-\frac{TS}{\omega}-\frac{\Omega J }{\omega}-\frac{\Phi Q}{\omega}\Bigr) \!\frac{\delta(L^{d-2}/G)}{L^{d-2}/G} -\frac{M}{\omega(d-2)}\frac{\delta (\omega L)^{d-2}}{(\omega L)^{d-2}},\quad \  
\end{align}
which, using \eqref{Ck}, \eqref{volomeg} 
and \eqref{xdictionary},  becomes  
\be
\delta \tilde E=\tilde T\delta S+\tilde \Omega \delta J+\tilde \Phi \delta \tilde Q+\mu \delta C-p\delta {\cal V}\,,\label{cftflaw}
\ee
which is the CFT first law, where
\ba
\mu&=&\frac{1}{C}( \tilde E - \tilde TS-\tilde \Omega J-\tilde \Phi \tilde Q)\label{mudef} \\
p &=&  \frac{\tilde E}{(d-2){\cal V}} \label{pEoS}
\ea 
recovering \eqref{UCFT} in the latter equation. 
The duality between the AdS-bulk first law \eqref{Eq8} 
and CFT-boundary first law
\eqref{cftflaw}
is now clear.

Eq.~\eqref{mudef} is the Euler relation for holographic CFTs. 
It is dual to the bulk Smarr relation
\eqref{Vsmarr}  for charged AdS 
black holes \cite{Visser:2021eqk,Karch:2015rpa}.
It follows (on the CFT side) from the proportionality of the thermodynamic quantities with the central charge ($\tilde E, \tilde   S, \tilde  J, \tilde Q \!\propto \!C$),   and occurs   in the deconfined   phase (dual to an AdS black hole geometry).  Note that
\eqref{mudef} has no dimension-dependent factors whereas the bulk Smarr relation~\eqref{Vsmarr} does. To see how this works, 
we can write   $PV$  as
a partial derivative of the CFT energy $\tilde E$
\beq
\label{PVterm}
- 2 P V = L \left ( \frac{\partial M}{\partial L}\right)_{A,J,Q,G}\!\!\!\! = L \omega \left (\frac{\partial \tilde E}{\partial L}\right)_{A,J,Q,G}  
\eeq   
and since  $\tilde E = \tilde E (S(A,G),J,\tilde Q(Q,L,G),C(L,G),V(L,\omega))$
as a function of bulk quantities, 
\eqref{Ck}, \eqref{volomeg} 
and \eqref{xdictionary} yield
\begin{align}
\label{partial}
&\left ( \frac{\partial \tilde E}{\partial L}\right)_{A,J,Q,G} \!\!\!\!\!\!=\frac{1}{L} ( \tilde \Phi \tilde Q + (d-2)\mu C - (d-2) p {\cal V})  \\
&\qquad \qquad \,\, =\frac{1}{L} ( (d-3) (\tilde E - \tilde \Phi \tilde Q) - (d-2) (\tilde \Omega J + \tilde T S))\,, \nonumber
 \end{align}
 using~\eqref{mudef} and
 \eqref{pEoS}. Inserting 
 \eqref{partial} into 
 \eqref{PVterm} yields 
 the bulk Smarr relation
 \eqref{Vsmarr}.

Eq.~\eqref{pEoS} is the CFT 
equation of state \eqref{UCFT}, derivable from CFT scaling symmetry.  Note that there is no  $p\mathcal V$ term in the Euler relation~\eqref{mudef}, reflecting the fact that  the internal energy is not an extensive variable on compact spaces at finite temperature in the deconfined phase.  This is a feature of holographic CFTs.  For 
$\omega L \tilde T  \gg 1$, namely  the high-temperature or large-volume regime, the $\mu C$ term becomes equal to $ - p \mathcal V$, and   the energy becomes extensive.   

 The rescaling property that yielded the Euler relation and equation of state  can be used to eliminate some of the terms in the first law 
 \eqref{cftflaw}. For example the $p\delta V$ term can be removed by 
rescaling
\ba
\label{re1}
\hat E&=&\omega L \tilde E \qquad \hat T=\omega L \tilde T \qquad \hat \Omega=\omega L \tilde \Omega \qquad   
 \hat \Phi = \omega L \tilde \Phi \qquad \hat \mu=\omega L \mu 
\label{eq14}
\ea
yielding
\ba
\delta \hat E&=&\hat T\delta S+\hat \Omega \delta J+\hat \Phi \delta \tilde Q+\hat \mu \delta C\,,\label{j4} \\
\hat E&=&\hat T S+\hat \Omega J+\hat \Phi \tilde Q+\hat \mu C\,,  \label{j5}
\ea
from \eqref{cftflaw} and
\eqref{mudef} respectively. Since 
 all thermodynamic quantities are now scale invariant,  the thermal description respects the symmetries of the CFT. 

Alternatively, the $\mu \delta C$ term can be eliminated from \eqref{cftflaw} using
\eqref{mudef}, giving
\ba
\delta \bar{E}&=&\tilde {T}\delta \bar{S}+\tilde {\Omega} \delta \bar{J}+\tilde {\Phi} \delta \bar{Q}-\bar{p} \delta {\cal V}\,,\label{j8}\\
\bar E&=& (d-2) \bar p {\cal V}\,, 
\ea
with the rescaled quantities: 
\be 
\label{re2}
\bar{E}=\frac{\tilde E}{C}\,,\quad \!
\bar{S}=\frac{S}{C}\,,\quad \!
\bar{J}=\frac{J}{C}\,,\quad \!\bar{Q}=\frac{\tilde Q}{C}\,,\quad \bar{p}=\frac{p}{C}\,.
\ee
Note that these barred quantities  are no longer proportional to $C$. All thermodynamic quantities now retain their standard dimensionality, and the `standard' thermodynamic first law, with $\bar{E}$ interpreted as internal energy, is recovered.

The laws  \eqref{j4} and \eqref{j5} were recently proposed in another context called
the  ``restricted phase space"   approach
\cite{Zeyuan:2021uol}, in which the  CFT volume ${\cal V}$ is kept fixed and variations in $C$ arise from variations of $G$ in the bulk.  The physical interpretation
is thus very different and the resultant holographic thermodynamics  has nothing to do with the original black hole chemistry. In contrast to the  restricted phase space approach, here variation of $C$  
(and of ${\cal V}$) 
is induced by variation of $\Lambda$ in the bulk,
with $G$ fixed
\cite{Ahmed:2023snm}.

\section{Holographic Thermodyamics for Charged AdS Black Holes}

We can now make use of the preceding considerations to analyze the thermodynamic behaviour of charged AdS black holes from a holographic viewpoint \cite{Cong:2021jgb}.
There are three pairs of conjugate quantities --     $(\tilde \Phi, \tilde Q)$, $(p, \cal V)$ and $(\mu, C)$ -- and so there are  eight   possible    thermodynamic (grand) canonical ensembles in the CFT.  Only three of these -- fixed $(\tilde Q, {\cal V}, C)$, fixed $(\tilde \Phi, {\cal V},C)$, and fixed $(\tilde Q, {\cal V}, \mu)$ --
  exhibit interesting phase behaviour or critical phenomena \cite{Cong:2021jgb}. 
  
 The action for electromagnetism minimally coupled to Einstein gravity is given by \eqref{LovMax}
 with $\alpha_{(k)}=0$ for $k\geq 2$. For later convenience we write this as
 \begin{equation} \label{EMaction}
 	I = \frac{1}{16\pi G} \int d^{d } x\sqrt{-g} \left (  R - 2 \Lambda - F^2  \right)
 \end{equation} 
 in $d$ spacetime dimensions, with  $\Lambda$ is given
in \eqref{P}. Note     the 
normalisation $1/16\pi G$  of the matter   action, which is  not standard.

The field equations yield the solution
 \begin{equation} \label{eq:metric1}
 	ds^2= -f(r)dt^2 + \frac{dr^2}{f(r)} + r^2 d \Omega_{d-2}^2
 \end{equation}
 in the spherically symmetric case,  where 
 \begin{equation} \label{eq:blackening}
 	f(r) = 1 + \frac{r^2}{L^2} - \frac{m}{r^{d-3}} + \frac{q^2}{r^{2d-6}} \, . 
 \end{equation}
 and $m$ is the mass parameter of the black hole. 
 It is related to the ADM mass by
  \begin{equation} \label{eq:admmass}
 	M= \frac{ (d-2)\Omega_{d-2}  }{16 \pi G}  m  
 \end{equation}
where $\Omega_{d-2}$ is area of the unit $d-2$ round sphere. The associated gauge potential is
  \begin{equation}
 	A = \left ( -\frac{1}{\zeta} \frac{q}{r^{d-3}} + \Phi\right) dt 
 \end{equation} 
and the electric charge is related to the charge parameter $q$ via
  \begin{equation} \label{eq:electriccharge1}
 	Q = \frac{(d-2)\Omega_{d-2}}{8\pi G} \zeta  \,  q \,, \qquad \text{with} \qquad \zeta =\sqrt{ \frac{2(d-3)}{d-2} } 
 \end{equation}
where $\Phi$ is  a constant  that plays the role of the electric potential.

Since $f(r_h)=0$ where $r_h$ is the (outer) horizon radius, we can write
  \begin{equation}
    m=   r_h^{d-3}\left (1 + \frac{r_h^2}{L^2} + \frac{q^2}{r_h^{2d -6}}    \right)
 \end{equation}
and  
 \begin{equation} \label{eq:electricpotential1}
 	\Phi = \frac{1}{\zeta}\frac{q}{  r_h^{d-3}} 
 \end{equation}
by  requiring $A_t(r_h)=0$. This choice implies that   $\Phi$ is the potential difference between the outer horizon and infinity. 
The    Hawking temperature $T$ and the Bekenstein--Hawking entropy $S$ 
are 
 \begin{equation} \label{eq:entropytemp}
    	T   = \frac{d-3}{4 \pi r_h} \left (1 + \frac{d-1}{d-3} \frac{r_h^2}{L^2} - \frac{q^2}{r_h^{2d  -6}}\right) \qquad   S  = \frac{\Omega_{d-2} r_h^{d-2}}{4G}  
 \end{equation}
which can be obtained by   computing the surface gravity 
defined with respect to the time translation Killing vector $\xi=\partial_t$ and the horizon area respectively.
  
It is straightforward to show that the thermodynamic quantities
\eqref{eq:admmass}, \eqref{eq:electriccharge1},
\eqref{eq:electricpotential1}, and 
\eqref{eq:entropytemp}
obey the Smarr relation
\eqref{Vsmarr} and first law
\begin{equation} \label{eq:extendedfirstlaw}
 	d M = T dS + \Phi dQ -\frac{V}{8\pi G_N} d \Lambda - \left ( M   - \Phi Q \right)\frac{dG}{G} 
 \end{equation}
whose latter term differs from
that in 
\eqref{Bulk1} due to the 
choice of normalization in the matter part of the action
\eqref{EMaction}. The effect of this is to redefine the bulk charge and potential as $Q_b \to \sqrt{G}Q$ and 
$\Phi_b \to \Phi/ \sqrt{G}$,
where $Q_b$ and $\Phi_b$ denote the charge and potential in previous sections, yielding
$\Phi d Q + \Phi Q dG / G = \Phi_b dQ_b + \frac{1}{2}\Phi_b Q_b dG/G$,  whose insertion into 
\eqref{eq:extendedfirstlaw} recovers \eqref{Bulk1}.

 The considerations of the previous section indicate that from \eqref{volomeg}
we can choose  the spatial volume of the CFT to be
 \begin{equation}
 \label{vol}
 	{\cal V} = \Omega_{d-2} R^{d-2}
 \end{equation}
by taking 
the boundary curvature radius $R$ to differ from $L$. ${\cal V}$  is the volume of a $(d-2)$-dimensional sphere of radius $R$ in the  CFT boundary geometry $\mathbb R \times S^{d-2}$, which corresponds to an ``area'' in the $d$-dimensional AdS bulk geometry. Asymptotically, 
 
  \begin{equation}
 	ds^2 = - \frac{r^2}{L^2 } dt^2 + \frac{L^2}{r^2}dr^2 + r^2 d \Omega_{d-2}^2 
\to 
ds^2 = - \frac{R^2}{L^2} dt^2 + R^2 d\Omega_{d-2}^2
\label{eq:cftmetric}
 \end{equation}
by choosing the Weyl factor relating the AdS and CFT metrics 
to be  $\lambda = R/r$. The holographic dictionary is then
 \cite{Chamblin:1999tk,Karch:2015rpa,Visser:2021eqk}
 \begin{equation} \label{eq:dictionaryentropyetc}
  S= \frac{A}{4G_N}\,, \qquad 	E = M \frac{L}{R} \,,\qquad  T = \frac{\kappa}{2\pi} \frac{L}{R} \,,\qquad \tilde \Phi =\frac{\Phi}{L}\frac{L}{R}\,,  \qquad \tilde Q = Q L \,.
 \end{equation}
for this choice of CFT metric.  
The $L/R$ factor appears because the bulk Schwarzschild time $t$ differs from  the boundary CFT time in \eqref{eq:cftmetric} by a factor $R/L$ \cite{Savonije:2001nd}.
The holographic dictionary  \eqref{vol} and \eqref{eq:dictionaryentropyetc} then implies (setting $J=0$) the first law 
\eqref{FirstHol} along with
the Smarr relation \eqref{mudef}
and equation of state \eqref{pEoS} for the CFT.

In analyzing the phase behaviour in the CFT, it is useful to write \cite{Dolan:2016jjc}
   \begin{equation}
 	x \equiv \frac{r_h}{L} \,,\qquad   y \equiv \frac{q}{L^{d-2}}\,.
 \end{equation}
upon which \eqref{eq:admmass}, \eqref{eq:entropytemp}, and \eqref{eq:dictionaryentropyetc} become
\begin{eqnarray}
\label{energytempcft}
E = \frac{d-1}{R} C x^{d-2} \left ( 1 + x^2 + \frac{y^2}{x^{2d-4}} \right) &\qquad& T = \frac{d-2}{4 \pi R} \frac{1}{x} \left ( 1 + \frac{d}{d-2} x^2 - \frac{y^2}{x^{2d-4}} \right)\\\
 \label{SQphi}
 	S = 4 \pi C x^{d-1} \qquad \tilde Q = 2 \alpha (d-1)  C  y  &\qquad& \tilde \Phi = \frac{1}{\alpha R} \frac{y}{  x^{d-2}} 
 \end{eqnarray}
and we obtain
\begin{equation}
 \label{mu}
 	\mu = \frac{x^{d-2}}{R} \left ( 1 - x^2 - \frac{y^2}{x^{2d -4}}\right)
 \end{equation}
for \eqref{mudef}.

\subsection{Thermodynamic Stability}

In analyzing various ensembles, the key quantity of interest is the free energy. The expression for the free energy (or thermodynamic potential) depends on the choice of ensemble, but in all cases the equilibrium state corresponds to its global minimum.  The state yielding the global minimum of the free energy is generally the thermodynamically stable state.

The sign of the specific heat (or heat capacity)
\be
C_P\equiv C_{P,J_1,\dots, J_N,Q_1,\dots,Q_n}=T\Bigl(\frac{\partial S}{\partial T}\Bigr)_{P,J_1,\dots,J_N,Q_1,\dots, Q_n}
\ee
provides information about local thermodynamic stability; if $C_P > 0$ then a given state is locally thermodynamically stable, though it may not be the most stable thermodynamic state. Considering the free energy as a function of temperature, locally stable states are those for which the 2nd derivative is negative, whereas locally unstable states have positive 2nd derivative.  
 
In what follows we shall find thermodynamically stable and unstable phases for the various ensembles, commenting on these in turn.

 \subsection{Fixed ($\tilde Q, {\cal V}, C$)}\label{sec:firstensemble}
 
 \begin{figure}
    \centering
    \includegraphics[scale=0.75]{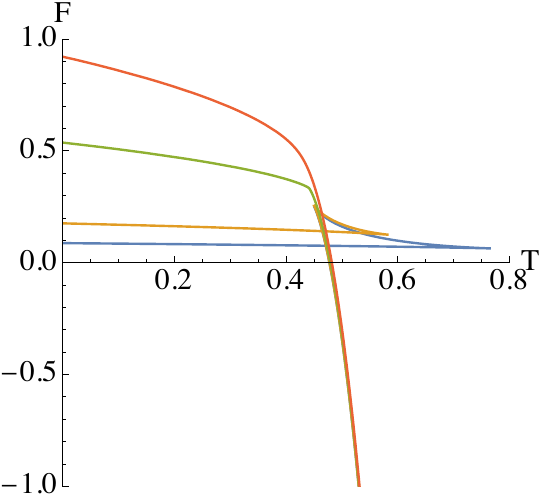} \hspace{1 cm}
    \includegraphics[scale=0.75]{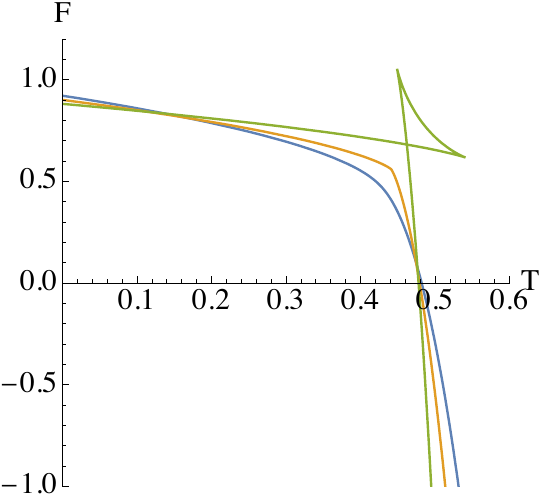}
    \caption{Free energy ${\cal F}$ vs. temperature $T$ plot   in $d=4$ for fixed $(\tilde Q,{\cal V},C)$, setting $R=1$. \textbf{Left}: 
    Plots for fixed $C=1$ for 
     $\tilde Q=0.1,0.2, 4/3 \sqrt{5}, 1$ (blue, orange, green, red). First-order phase transitions occur for $ \tilde Q<  \tilde Q_{crit}$ (blue, orange) as the    ``swallowtail'' 
     structure indicates, whereas 
there is a second-order phase transition for $\tilde Q= \tilde Q_{crit}$ (green). No phase transitions occur  for $\tilde Q  > \tilde Q _{crit}$ (red). 
    \textbf{Right}: Plots for fixed $\tilde Q = 1$ for 
      $C=1, 3\sqrt{5}/4, 4$ (blue, orange, green). 
There is a critical central charge above which a  swallowtail
is present (green) and there is
a first-order phase transition. For $C<C_{crit}$ (blue)
no phase transitions take place, and for   $C=C_{crit}$ (orange) 
there is a second order phase transtion.  The  apparent triple intersections of the  blue, orange and green curves is a consequence of plotting resolution.
This figure originally appeared in ref.~\cite{Cong:2021jgb}.
    }
    \label{fig:qvcFT}
\end{figure}
 
Fixing  the charge $\tilde Q$, spatial volume ${\cal V}$ and   central charge $C$ corresponds to the   canonical ensemble, whose      thermodynamic potential
 \begin{align}
 \label{F1}
 	{\cal F}  \equiv E - T S 
 	 = C \frac{x^{d-2}}{R} \left( 1 - x^2 + (2d-3) \frac{y^2}{x^{2d-4}}\right)  
 \end{align}
  is the  Helmholtz free energy. Since
  \begin{equation}
     d{\cal F}  = dE-TdS-SdT = -SdT+\tilde \Phi d\tilde Q-pd{\cal V}+\mu dC
 \end{equation}
 ${\cal F} $ is    stationary at   fixed   $(T,\tilde Q,{\cal V},C)$.

To understand phase behaviour, we wish to plot ${\cal F}$ as a function of $T$, which can both be regarded as functions of ($\tilde Q,R,C,x$); from \eqref{vol}, fixing 
$R$ is equivalent  to fixing ${\cal V}$.  
Using \eqref{SQphi} to eliminate $y$ yields
 
 \begin{equation}
 \begin{aligned}
 	{\cal F}   &=   C \frac{x^{d-2}}{R} \left( 1 - x^2 +  \frac{2d-3}{4 \alpha^2 (d-1)^2 C^2}  \frac{{\tilde Q}^2}{x^{2d-4}}\right)\,,\\
 	T &= \frac{d-2}{4 \pi R} \frac{1}{x} \left ( 1 + \frac{d}{d-2} x^2 -  \frac{2d-3}{4 \alpha^2 (d-1)^2 C^2}  \frac{\tilde Q^2 }{x^{2d-4}} \right)\,.
 \end{aligned}
 \end{equation}
 and  plotting  ${\cal F} $ in terms of $T$ is straightforward. Note that dependence on    $R$ is trivially fixed by scale invariance
 and so $R=1$ can be taken without loss of generality.
  
The results are illustrated in  
Figure \ref{fig:qvcFT}
for a variety of 
choices of $\tilde Q$ relative to fixed $C$ (left) and of $C$ relative to fixed $\tilde Q$ (right).
  In both cases the free energy displays ``swallowtail'' behaviour connecting three different branches, but in the former case this only occurs   for $\tilde Q< \tilde Q_{crit}$ and in the latter case for $C > C_{crit}$, where $\tilde Q_{crit} $ and $C_{crit}$ are  critical values whose ratio is   \cite{Cong:2021jgb}
\begin{equation}
 \label{Ccrit}
 	\frac{C_{crit}}{\tilde Q_{crit}} =\frac{1}{2 \alpha (d-1) y_{crit}  }   = \sqrt{\frac{ (2d-3)}{8(d-2)  }}\frac{1}{x_{crit}^{d-2}} \, .
 \end{equation}
 In Figure \ref{fig:qvcFT}  the former case is shown on the left and the latter case is shown on the right. 

We see the appearance of a both a critical electric charge (left panel) and a critical central charge (right panel). If $C$ is fixed, for $\tilde Q > \tilde Q_{crit}$
there are no phase transitions, whereas for $\tilde Q < \tilde Q_{crit}$   a swallowtail appears, indicating a first-order phase transition.   Since the CFT entropy    $S=4 \pi C x^{d-1}$,  small AdS charged black holes  are dual to CFT thermal states with small $S/C$, namely  states with low entropy per degree of freedom. Such states occur for
low temperatures and have the smallest ${\cal F}$.  As $T$ increases, 
the self-intersection point of the curve is eventually reached, and there is then a transition to 
a CFT state  of  high entropy (per degree of freedom),  corresponding to large AdS black holes.  These states now have the lowest free energy and are along the steep (near-vertical)  branch of the curve. As $\tilde Q$ increases, the phase transition occurs at larger values of $T$, whilst  the swallowtail shrinks in size, vanishing at  $\tilde Q=\tilde Q_{crit}$, at which point the phase transition   becomes second order. This 
behaviour   is commensurate with the canonical ensemble for AdS black holes at fixed charge~\cite{Chamblin:1999tk,Kubiznak:2012wp}; the critical point $(\tilde Q_{crit}, T_{crit})$  depends on the value of $C$.

There is likewise for any fixed 
$\tilde Q$ a critical value of $C$
{\it above} which there is a first order phase transition, as shown in the right panel of 
figure~\ref{fig:qvcFT} \cite{Cong:2021fnf}. For $C>C_{crit}$ the   transition  is again between states with low- and high-entropy per degree of freedom, and at  at $C_{crit}$ there is a second-order phase transition;  a single phase is present for $C<C_{crit}$. As 
$C$ decreases 
the value of ${\cal F}$ at which the first-order phase transition occurs also decreases  whereas
for fixed $C$, ${\cal F}$
increases  as $\tilde Q$ decreases.

 \subsection{Fixed ($\tilde\Phi, {\cal V}, C$)}
 
  \begin{figure}
     \centering
     \includegraphics[scale=0.75]{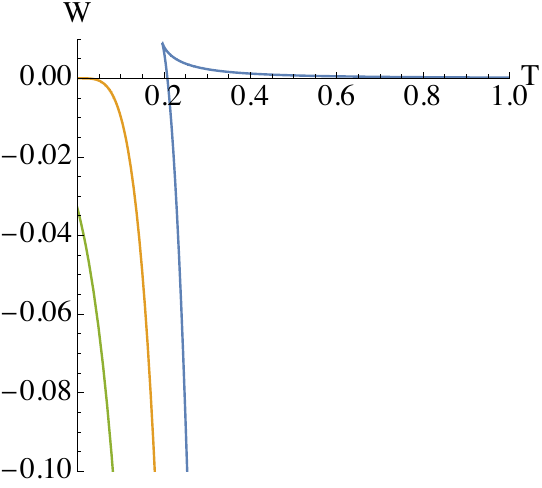} \hspace{1cm}
     \includegraphics[scale=0.75]{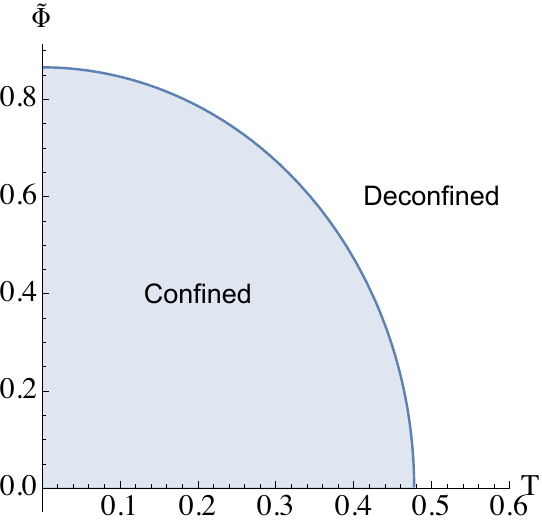}
     \caption{Free energy ${\cal W}$ vs. temperature $T$ plot and phase diagrams in $d=4$ for   fixed $(\tilde \Phi,{\cal V},C)$, setting $R=1$. \textbf{Left}: Plots for the parameters  $C=1$ and $\tilde\Phi=0.9\Phi_c$ (blue), $\tilde\Phi=\Phi_c =  \sqrt{3}/2$ (orange) and $\tilde\Phi=1.1\Phi_c$ (green). For $\tilde\Phi<\Phi_c$  the free energy curve  is 
      two branches joining in a cusp, with the upper/lower branches  corresponding
      to a low/high-entropy states.  Now there is a   first-order phase transition at ${\cal W}=0$ between the high-entropy ``deconfined'' state and a ``confined'' state,  dual to a generalised Hawking-Page   transition between a large black hole and thermal AdS,  depending on the value of $\tilde \Phi$. 
     For $\Phi \ge \Phi_c$  the curve never has ${\cal W}>0$ and so no phase transition occurs. \textbf{Right}: Phase diagram
     for $C=1$; the  coexistence curve is a line of (de)confinement phase transitions in the CFT. At
     $T=0$ the transition takes place at $\tilde \Phi =\tilde \Phi_c$ whereas it occurs at the Hawking-Page temperature $T=T_c=3/2\pi$ for  $\tilde \Phi=0$. This figure originally appeared in ref.~\cite{Cong:2021jgb}.} 
     \label{fig:WTplot}
 \end{figure} 
 
Fixing the potential $\tilde \Phi$, instead of the charge $\tilde Q$, yields
  \begin{equation} \label{eq:freeenergy2}
 	{\cal W} \equiv E - TS - \tilde \Phi \tilde Q = \mu C =C \frac{x^{d-2}}{R} \left ( 1 - x^2 - \frac{y^2}{x^{2d -4}}\right)  
 \end{equation}
for the thermodynamic potential of this ensemble, which is the    Gibbs  free energy.  The middle equation follows from \eqref{mudef}, and so the free energy ${\cal W} \propto \mu$.  The last equality follows from \eqref{mu}. 

Writing   $\tilde \Phi = y/ \alpha R x^{d-2}$ yields
 \begin{equation}
 \label{GibssW}
 	{\cal W}   =C \frac{x^{d-2}}{R} \left ( 1 - x^2 -  \alpha^2 R^2 \tilde \Phi^2\right)  \qquad
  T = \frac{d-2}{4 \pi R} \frac{1}{x} \left ( 1 + \frac{d}{d-2} x^2 - \alpha^2 R^2 \tilde \Phi^2 \right)
 \end{equation}
 allowing a parametric plot of  ${\cal W}(T)$ using $x$ as parameter, 
 shown in Figure \ref{fig:WTplot}.

 In this case at ${\cal W}=0$ there is a   first-order phase transition  between the high-entropy ``deconfined'' state and a ``confined'' state. This is  dual to a generalised Hawking-Page   transition between a large black hole and thermal AdS,  depending on the value of $\tilde \Phi$. 
Setting ${\cal W}=0$ and eliminating $x$ in terms of $T$ yields
\begin{align}
 \tilde \Phi &= \frac{\tilde \Phi_c}{T_c} \sqrt{T_c^2 - T^2} \,, \qquad \text{with} \qquad T_c = \frac{d-1}{2\pi R} \,,\qquad \tilde \Phi_c =\frac{1}{\alpha R} =\frac{1}{R} \sqrt{\frac{d-1}{2d-4} }
 \label{linefixedphi}
 \end{align}
which describes the  coexistence line between the two phases shown in the right panel of figure~\ref{fig:WTplot}.  At  $T=0$ the phase transition  occurs at $\tilde \Phi = \tilde \Phi_c$, and at $\tilde \Phi=0$ the   transition is  at $T=T_c$, and is equivalent to the standard Hawking-Page phase transition in the bulk.

For $\tilde\Phi<\tilde \Phi_c$ (blue curve in the left panel of
figure~\ref{fig:WTplot}) the   free energy curve is a cusp.
The upper branch corresponds to low-entropy CFT states (dual to small black holes) whereas the lower branch corresponds to  high-entropy CFT states  (dual to large black holes). The temperature as a function of $x$ is minimized   at the cusp 
\begin{equation}
\begin{aligned}
   &\left ( \frac{\partial T}{\partial x} \right)_{\tilde \Phi}=0 \quad \text{at}\quad x_{cusp} = \sqrt{\frac{d-2}{d}(1-\alpha^2  R^2\tilde \Phi^2)} \,,\\
   & \Rightarrow \qquad T_{cusp} = \frac{1}{2\pi R}\sqrt{d(d-2)(1-\alpha^2  R^2\tilde \Phi^2)}  \label{cusp}
   \end{aligned}
\end{equation}
which is turn is the coldest possible (unstable) deconfined state.
The specific heat is positive/negative on the  lower/upper branch indicative of  stable/unstable solutions.

 \subsection{Fixed ($\tilde Q$, $ {\cal V}$, $\mu$)}
 \label{sec:thirdensemble}
 
 \begin{figure}
     \centering 
     \includegraphics[scale=0.75]{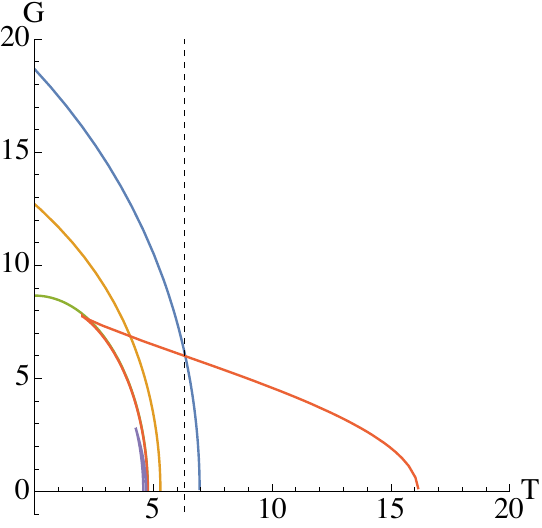} \hspace{1cm}
     \includegraphics[scale=0.335]{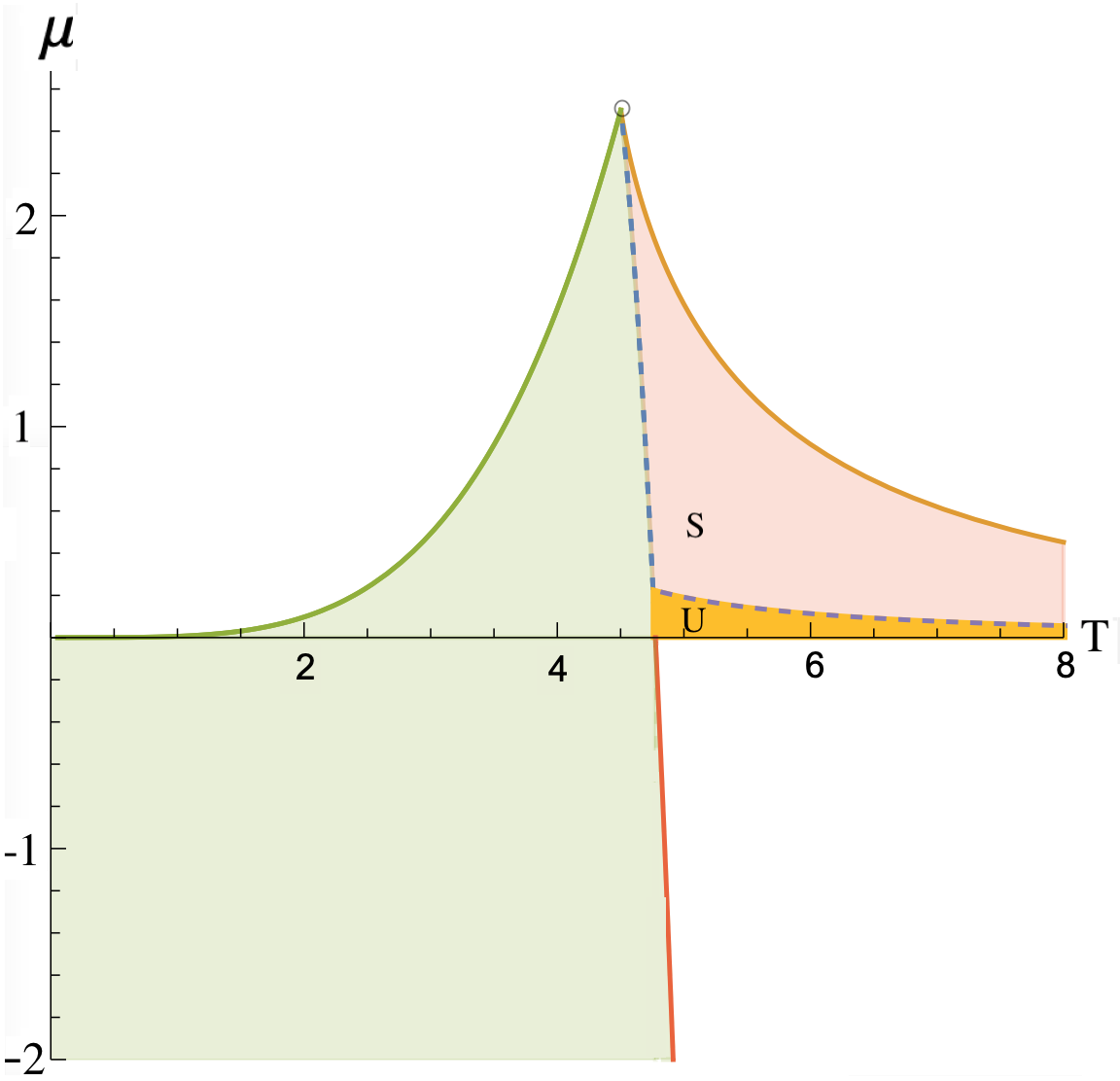}
     \caption{Free energy ${\cal G}$ vs. temperature $T$ plot and phase diagram for   fixed $(\tilde Q,{\cal V},\mu)$ 
     for $R=0.1$ and  $\tilde Q=1$ in $d=4$. 
      \textbf{Left}: Free energy ${\cal G}(T)$  for  $\mu=-60,-10,0,1/10,2$ (blue, orange, green, red, purple curves). We see the existence of a  single stable phase for $\mu\leq 0$ 
      (blue, orange, green curves). There are branches for
       $\mu>0$ (red, purple curves) that meet at a cusp $T=T_{0}$ \eqref{cusptemp}, corresponding to        two CFT phases. These branches meet the ${\cal G}=0$ line at the temperatures $T_1\leq T_2$ \eqref{TLR}. The upper/lower  branch corresponds to a low/high entropy state.  The specific heat is positive 
       along the upper branch near and below $T=T_2$, but becomes negative at some intermediate temperature $T_{int}$, indicated by the black dashed line for the red curve. Nevertheless, this phase is the thermodynamically preferred phase for $T > T-1$.        Once $T=T_1$, there is a   a zeroth-order phase transition to the high-entropy CFT phase given by the lower branch. As $T$ further decreases this phase is preferred, until $T=T_0$ at which point both phases coincide. Below this temperature there are no CFT states. 
       \textbf{Right}: The  $\mu-T$ phase diagram indicating the different phases. High-entropy phases are in the   green shaded region bounded by the green curve. Here the heat capacity is positive. The orange and red  regions on the right above the $T$-axis correspond 
       respectively to stable  (``S'')
         and unstable (``U'')   low-entropy phases having respective  positive and negative heat capacities. 
       White regions indicate that no solution exists. This figure originally appeared in ref.~\cite{Cong:2021jgb}. }
     \label{fig:GTplot}
 \end{figure}

For fixed chemical potential $\mu$, the free energy is
\begin{equation} \label{eq:freeenergy3}
	{\cal G} \equiv E - T S - \mu C = \tilde \Phi \tilde Q 
\end{equation}
where \eqref{mudef} was used to obtain the last equality. 
 The differential of ${\cal G}$ is
\begin{equation}
    d{\cal G} = - S dT + \tilde \Phi d \tilde Q - p d {\cal V} - C d \mu 
\end{equation}
using the first law \eqref{FirstHol}, and so ${\cal G}$ is  stationary at fixed $(T, \tilde Q, {\cal V}, \mu).$ 

Although fixing the chemical potential instead of the  number of degrees of freedom is very natural in   thermodynamics, this  ensemble has only recently been considered \cite{Cong:2021jgb}. Here fixed $\mu$ implies that the central charge can vary,  which is only possible if we consider a family of holographic CFTs having  different central charges.  In the gravity theory this corresponds to allowing for variations of $\Lambda$ (and ${\cal G}$).  

Using  \eqref{mu} and \eqref{SQphi} gives
\begin{align}
 {\cal G}    &= \frac{|\tilde Q|}{\alpha  R  } \sqrt{1-x^{ 2}- \frac{R\mu}{x^{d-2}} } \,,	\label{TG}\\ \qquad 	T &= \frac{d-2}{4 \pi R} \left (\frac{R\mu}{x^{d-1}}  + \frac{2(d-1)}{d-2}x \right) \,. \label{TG2}
\end{align}
where \eqref{eq:freeenergy3} implies that
$$
{\cal G}=\frac{L}{R} \Phi Q=\frac{L}{R} \frac{(d-1)\Omega_{d-1}}{8\pi G} \frac{q^2}{r_h^{d-2}} \geq 0
$$
which is why the  absolute value of $\tilde Q$ appears
in \eqref{TG}.
 
In figure~\ref{fig:GTplot}
a plot of ${\cal G}(T)$ is given for several values of $\mu$, along with the phase diagram for this ensemble. 
For $0<\mu<\mu_{coin}$ (red and purple curves) there are two branches, cutting the ${\cal G}=0$ line at two temperatures 
\begin{equation} \label{TLR}
    T_1=\frac{3-4 \mu  R+3 \sqrt{1-4 \mu  R}}{\sqrt{2} \pi  R \left(1+\sqrt{1-4 \mu  R}\right)^{3/2}}\,,\qquad T_2=\frac{3-4 \mu  R-3 \sqrt{1-4 \mu  R}}{\sqrt{2} \pi  R \left(1-\sqrt{1-4 \mu  R}\right)^{3/2}}
\end{equation}
for $d=4$, with $T_2\ge T_1$. For any value of $d$ the two intersection points  correspond  to the two positive roots of the function  $f(x)=1-x^{ 2}- \frac{R\mu}{x^{d-2}}$ in  \eqref{TG}. The lower  branch is a high-entropy state, and 
is  thermodynamically preferred for $T_{0}<T<T_1$, where 
\begin{equation} \label{cusptemp}
    T_{0} = \frac{d}{2\pi R} \left ( \frac{d-2}{2} \mu R\right)^{1/d}
\end{equation}
is the   temperature at the cusp, obtained from solving
$\frac{\partial T}{\partial x}|_{\mu} = 0$.
For larger values of $T$ this phase does not exist, and the CFT undergoes a zeroth-order phase transition to the low-entropy CFT phase on the upper branch. This    branch has positive specific heat near and below $T=T_2$, but  becomes negative at some intermediate temperature 
\be
T_{int}=\frac{(d-2) \mu  \left(\frac{2}{d \mu  R}\right)^{\frac{d-1}{d-2}}}{4\pi }+\frac{ (d-1) \left(\frac{2}{d \mu
    R}\right)^{-\frac{1}{d-2}}}{2\pi  R}, \label{intermediatetemp}
\ee
indicated (for the red curve) by the black dashed line in the left panel of figure~\ref{fig:GTplot}. 
For $T_{int} > T > T_0$  the low-entropy phase has  negative heat capacity and so is   unstable for these low temperatures. The   temperatures $T_{int}$ and   $T_1$, at which the zeroth-order phase transition occurs, coincide for two values of the chemical potential. One is 
at $\mu=\mu_*$, which 
is at
\be 
 \mu_*=\frac{1}{324R} \left(16 \sqrt{7}-35\right)
 \qquad 
  T_*= \frac{2 }{3 \pi R}\sqrt{2+8/\sqrt{7}}
 \ee
 for $d=4$.   
This means that 
for $0<\mu <\mu_*$  the phase transition happens between the \emph{unstable} low-entropy state and  the stable high-entropy state. For larger values of $\mu$  the zeroth-order phase transition takes place between the \emph{stable} low-entropy phase and the high-entropy phase. This occurs in the region
$ \mu_* <\mu<\mu_{coin}$, where
$\mu_{coin} = \frac{2}{dR} \left ( \frac{d-2}{d}\right)^{\frac{d-2}{2}}$ is the value of $\mu$ at which  
$T_2=T_1=T_{coin} = \frac{\sqrt{d(d-2)}}{2 \pi R} 
=T_0(\mu_{coin})$. 
The free energy  is not real for $\mu > \mu_{coin}$, and no CFT phases exist for
these values of $\mu$.

As $\mu$ decreases, $T_2$ becomes increasingly large and $T_1$ decreases. For  
 $\mu=0$,   $T_2\rightarrow \infty$ and
 $T_1=\frac{3}{2\pi R} $. A qualitative change takes place at $\mu=0$ (green curve) -- the free energy curve becomes single-valued (and monotonically decreasing). This persists for all values of $\mu < 0$ (blue, orange curves). The single-valuedness indicates a single phase in the CFT.  
 
Although the zeroth-order phase transition can take place as temperature increases above $T_1$, such a transition has the peculiar feature that the entropy decreases during the process, in contradiction to the second law of thermodynamics. The phase transition from the low-entropy to the high-entropy state happens for $T$  decreasing below $T=T_1$, increasing the entropy   during the process. Such a transition
 seems  more favourable  thermodynamically.

The phase diagram in the right panel of figure~\ref{fig:GTplot}
shows the high-entropy phase in green and the low-entropy phase in red (for the stable region having positive heat capacity) and orange (for the unstable region with negative heat capacity). The coexistence curve between the high and low entropy phases is the boundary between the green region and the red/orange one.
 There are no solutions in the white regions.  The boundary of the green high entropy region is determined by $T_1$ and is indicated by the red line.
The right and left boundaries
for $\mu>0$ are respectively set by $T_2$ and $T_0$; these meet 
   at the coincidence point,  $\mu =\mu_{coin}$.  The boundary between the red (stable) and orange (unstable) low-entropy phases is   set by  $T_{int}$. The two coexistence lines (the intersection of the two dashed lines) meet at  the point $(\mu_*,T_*).$

\subsection{Other  ensembles and criticality}
\label{sec:otherensembles}

Critical points in       the $\tilde Q - \tilde \Phi$ and $C- \mu$ planes can both be shown to have   mean field critical exponents. In the former case
criticality has been observed   \cite{Dolan:2016jjc} but the latter case is a genuinely novel recent discovery \cite{Cong:2021jgb}.  
None of the other five ensembles 
 display critical phenomena or phase transitions, though some ensembles have two different branches corresponding to low and high entropy phases
 \cite{Cong:2021jgb}.

\section{Conclusion}
\label{secConc}

Holographic Black Hole Chemistry is now on a considerably firmer footing than it was a few years ago. The duality between the first laws in
the AdS bulk \eqref{Eq8} and dual CFT
\eqref{cftflaw} is now clear, as is the duality between the bulk Smarr relation \eqref{Vsmarr} and the CFT Smarr relation \eqref{mudef}. The stage is set to obtain a new and deeper understanding of gauge-gravity duality from a thermodynamic perspective.

One possible line of research is to better understand going beyond the large-$N$ limit. This has been done to some extent for the holographic Smarr relation in section~\ref{sec2}
\cite{Sinamuli:2017rhp}. However in   $SU(N)$   gauge theories with conformal symmetry such as $\mathcal N=4$ supersymmetric Yang-Mills theory, the central charge $C\sim N^2$.  Changing the rank $N$ of the gauge group is tantamount to changing the original theory, and so variations of $C$ can be regarded as  moving within the space of theories and changing the number of degrees of freedom $N^2$. 
But this cannot be done continuously since   $N$ is an integer.  In the large-$N$ limit,  $\Delta N/ N \ll 1$, and hence 
$\mu dC \sim d C/C \ll 1$, which makes sense in this limit.  However subleading terms are induced by higher curvature corrections, as pointed out in section~\ref{sec2}, and understanding the variation of $C$ in this case merits further study.

  There have also been recent studies on rotating AdS black holes
\cite{Ahmed:2023dnh,Gong:2023ywu} where, as for charged black holes, the (inverse) central charge plays a role  similar to that of thermodynamic pressure. 
There are 8 possible ensembles, all of which have classical superradiant instabilities in the bulk; what this means for the CFT remains to be understood.  Other studies are emerging, including an investigation of the topological properties of bulk/boundary phase transitions \cite{Zhang:2023uay}, and
understanding the holographic CFT of accelerating black holes \cite{Arenas-Henriquez:2023hur}.  
Recent work on holographic complexity has shown that the 
the complexity of formation
scales with bulk thermodynamic volume (and not entropy) 
for large black holes, something particularly manifest for rotating black holes
\cite{AlBalushi:2020rqe,AlBalushi:2020heq,Bernamonti:2021jyu,Zhang:2022quy}.

These  results but are a small sample of what lies ahead.
The holographic implications for the full panoply of results  in Black Hole Chemistry \cite{Kubiznak:2016qmn} remain to be explored!

\acknowledgments

This work was supported in part by the Natural Sciences and Engineering Research Council of Canada. I am grateful to Moaathe Belhaj Ahmed, Wan Cong, David Kubiznak, Cedric Sinamuli, and Manus Visser for their collaborations that led to the results presented here.

\bibliography{main}

\end{document}